\documentclass{PoS}
\usepackage{tabularx}
\usepackage{float}
\usepackage{color}
\usepackage{float}
\usepackage{mathbbol}
\usepackage{amssymb}
\usepackage{amsmath}
\usepackage{wrapfig}
\usepackage{textcomp}

\title{Recent pulsar results from VERITAS on Geminga and the missing link binary pulsar PSR J1023+0038}

\ShortTitle{Geminga and PSR J1023+0038 results with VERITAS}

\author{\speaker{{G. T. Richards} for the VERITAS Collaboration\thanks{veritas.sao.arizona.edu}} \\Georgia 
Institute of Technology\\
        E-mail: \email{gtrichards@gatech.edu}}

\abstract{In recent years, the {\it Fermi}-LAT gamma-ray telescope has detected a population of over 160 gamma-ray 
pulsars, which has enabled the detailed study of electromagnetic radiation from pulsars at energies above 100 MeV. 
Further, since the surprising detection of the Crab pulsar in very high-energy (VHE; E > 100\,GeV) gamma rays by 
the MAGIC and VERITAS collaborations, there has been an ongoing effort in the astrophysics community to 
detect new pulsars in the VHE band. However, the Crab remains the only pulsar so far detected in VHE gamma rays, 
raising the question of whether or not the Crab is unique and also making it more difficult to constrain model 
predictions that attempt to explain the emission. Presented here are recent VERITAS results from observational 
campaigns on the brightest northern-hemisphere high-energy gamma-ray pulsar Geminga and the missing link binary 
pulsar PSR J1023+0038, which have both resulted in upper limits on a possible VHE flux. These limits are 
placed into context with the current theoretical framework attempting to explain the origin of the gamma-ray 
emission from pulsars. Additionally, future plans for pulsar observations with VERITAS will be briefly discussed.}

\FullConference{The 34th International Cosmic Ray Conference,\\
		30 July- 6 August, 2015\\
		The Hague, The Netherlands}

\begin{document}
\nocite{*}

\section{Introduction}

The field of pulsar astronomy has entered a new era in the time since the 
launch of the {\it Fermi} Large Area Telescope (LAT), which has now succeeded in detecting over 160 pulsars in the 
high-energy (HE; E > $100$\,MeV) gamma-ray 
band\footnote{\detokenize{https://confluence.slac.stanford.edu/display/GLAMCOG/Public+List+of+LAT-Detected+Gamma-R
ay+Pulsars/} } .  These detections have greatly facilitated the detailed study of electromagnetic radiation from 
pulsars at the highest energies.  The spectral energy distributions (SEDs) that have been seen for pulsars in gamma 
rays are all well-characterized by a broad curvature radiation component that originates due to the electrons and 
positrons that fill the magnetosphere and follow curved trajectories as they are confined to magnetic field lines.  
A curvature radiation component has a natural end where the emission becomes radiation-reaction 
limited~\cite{1986ApJ...300..522C}, resulting in a spectral break at a few 
GeV~\cite{2013ApJS..208...17A}.  The 
spectral ``tails'' of pulsars in the GeV band are well-described by an exponential cutoff as expected for 
curvature radiation~\cite{2013ApJS..208...17A}, though statistics are sparse above $\sim$10\,GeV due to the 
decreasing sensitivity of the {\it Fermi}-LAT at these high energies.  

In recent years, a few pulsars have been detected at energies considerably above the HE spectral break.  The Vela 
pulsar, the brightest steady gamma-ray source seen by the {\it Fermi}-LAT~\cite{2009ApJ...696.1084A}, has been 
detected from the ground by H.E.S.S.\footnote{\detokenize{http://www.mpg.de/8287998/velar_pulsar/}} above 
$30$\,GeV 
(Gajdus, M., these proceedings)~\cite{gajdus} 
and by the {\it Fermi}-LAT above $50$\,GeV~\cite{leung2014}.  The H.E.S.S. detection of the Vela pulsar makes it 
the second pulsar to have been detected in gamma rays from the ground.  The first, the Crab pulsar, has been 
detected in the VHE band above 100\,GeV by MAGIC~\cite{aleksic2011b} and VERITAS~\cite{aliu2011} and has posed 
a challenge in that the curvature radiation scenario is not adequate for a full explanation of the 
gamma-ray radiation; the combined {\it Fermi}-LAT and VERITAS SED favors a power-law fit above 
$\sim$10\,GeV~\cite{aliu2011}.  Current models that have been proposed to explain the VHE emission from the Crab 
pulsar include inverse-Compton (IC) scattering scenarios in the outer magnetosphere~\cite{lyutikovetal2012, 
du2012, lyutikov2012} and beyond the light cylinder~\cite{aharonian2012, petri2012}, and synchrotron radiation 
from magnetic reconnection events~\cite{mocholpetri2015}.  Whether or not the VHE emission from the Crab pulsar is 
exceptional with respect to the rest of the gamma-ray pulsar population remains to be seen.

In this paper we review recent results from the VERITAS collaboration on the Geminga pulsar and PSR J1023+0038. 
The paper is 
structured in the following way: in Sections 1.1 and 1.2 we review the Geminga pulsar and PSR J1023+0038, 
respectively; in Section 2 we discuss the VERITAS observations, analyses, and main results; and in Section 3 we 
present a brief overview of future pulsar science plans with VERITAS.


\subsection{The Geminga Pulsar}

Formally known as PSR J0633+1746, the Geminga pulsar is located at the relatively close distance of 
$\sim$200\,pc~\cite{caraveo1996, faherty2007} and is the second-brightest steady HE gamma-ray source in the sky.  
The Geminga pulsar has a spin period of $\sim$240\,ms and a spin-down luminosity of $3.2\times10^{34}$\,erg 
s$^{-1}$~\cite{bignamicaraveo1996}.  It is the first GeV-emitting pulsar detected with no known radio 
counterpart~\cite{1992Natur.357..306B} and was, in fact, first discovered in gamma rays by the {\it 
SAS-2}~\cite{fichtel1975} and {\it COS-B}~\cite{bennett1977} satellites.  Though radio searches have failed to 
reveal any hints of emission, the Geminga pulsar is detected in UV~\cite{kargaltsev2005} and 
X-rays~\cite{halpernholt1992}.  Similar to the Crab pulsar in the high-energy band, Geminga shows a double-peaked 
phaseogram separated by a ``bridge'' of emission above the background, with only one of the peaks remaining 
dominant above about 10\,GeV.  As has been seen for other gamma-ray pulsars, the spectrum of Geminga above 
100\,MeV 
is can be described by a power law with an exponential cut-off~\cite{2013ApJS..208...17A}, characteristic of 
the 
curvature radiation processes thought to dominate in the magnetosphere.  However,~\cite{lyutikov2012} reported 
that the spectrum is better described by a simple power-law above the break energy, bringing into question that 
the spectrum is best described by the exponential cut-off expected in a curvature radiation scenario. 


\subsection{PSR J1023+0038}

The system PSR J1023+0038 is an eclipsing binary system located at a distance of $1370\pm40$\,pc~\cite{deller2012} 
containing a millisecond pulsar (MSP) with a rapid rotation period of 1.69\,ms orbiting a 
non-degenerate companion star every $\sim$5\,hr, found in a survey in 2008 by the Green Bank 
Telescope~\cite{archibald2009}.  Before the discovery of the MSP, the system was first identified as a 
low-mass X-ray binary (LMXB) with an accretion disk~\cite{2005AJ....130..759T} in 2001.  Later, optical and X-ray 
observations revealed the disappearance of the accretion disk, implying a change to the MSP 
state~\cite{2004MNRAS.351.1015W}, though the MSP was not identified until 2008.  However, in 2013 June the radio 
pulsations disappeared~\cite{2013ATel.5513....1S}, and the system was shown to have reformed an accretion 
disk~\cite{2013ATel.5514....1H}, thus reverting back to the LMXB phase and constituting the first time a system 
has 
been seen to oscillate between the two states.  Furthermore, it has been thought that MSPs are old pulsars that 
are spun up to rapid rotation frequencies through the accretion of material from a companion star~\cite{alpar1982} 
in a process sometimes referred to as ``recycling,'' and the observed behavior of the so-called ``missing-link'' 
PSR J1023+0038 system has helped solidify the recycling scenario as the preferred explanantion for the origin 
rapid rotation of MSPs. To our knowledge, this work represents the first time that this intriguing system has 
probed for possible VHE gamma-ray emission.


\section{Observations, Analysis, and Results}

The VERITAS array of four 12\,m diameter imaging atmospheric Cherenkov telescopes is located at the Fred 
Lawrence Whipple Observatory (FLWO) in southern Arizona (31$^{\circ}$ 40\textquotesingle\,N, 110$^{\circ}$ 
57\textquotesingle\,W, 1.3\,km a.s.l.) and started full array 
operations in 2007.  The telescope reflectors each consist of 345 hexagonal mirror facets, and the cameras 
comprise 499 photomultiplier tubes giving a total field of view of $\sim$3.5$^{\circ}$.  VERITAS is sensitive to 
gamma-ray photons in the energy range $0.85$ to \textgreater\,$30$\,TeV with a sensitivity to detect a 1\% Crab 
Nebula source in 
$\sim$25 hr.  It has an energy resolution of $15$--$25$\%, an angular resolution of $0.1^{\circ}$ at 68\% 
containment, and a pointing accuracy error of less than 50 arcseconds~\cite{2008AIPC.1085..657H}.

\subsection{Geminga}
\begin{figure}[H]
\center
\includegraphics[width=5.0in]{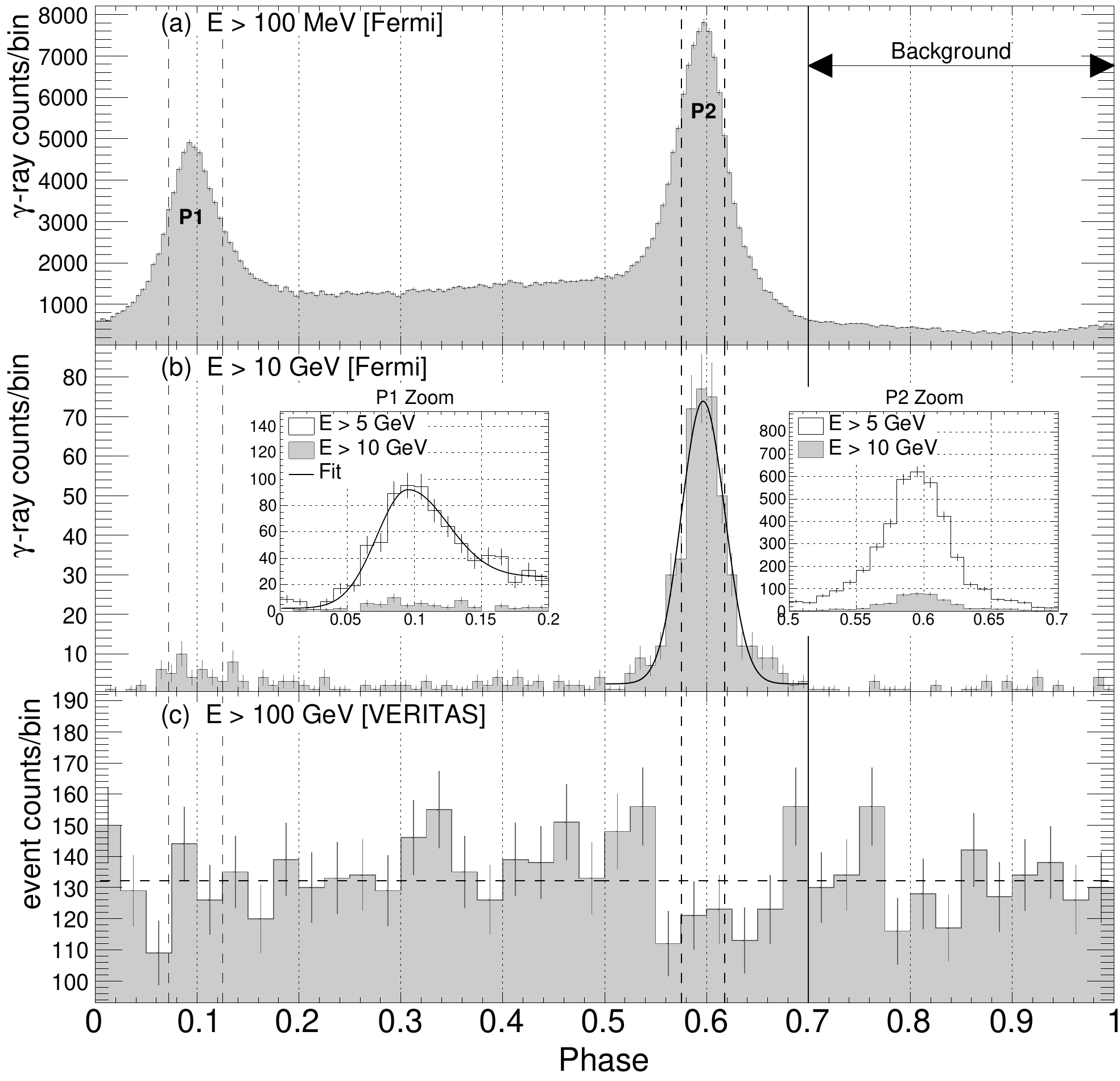}
\caption{Phase-folded light curves for the Geminga pulsar for both the {\it Fermi}-LAT (panels a and b) and 
VERITAS data (panel c).  In panel a, the phase-folded light curve of the {\it Fermi}-LAT data with an energy 
threshold of E $> 100$\,MeV is shown, and the region defined as background is indicated by the arrow.  Panel b 
shows the data with a higher energy threshold of 10\,GeV, and the asymmetric Gaussian fit to P2 is represented by 
the solid black curve.  The inset zoom panels for P1 and P2 show the respective peaks with energy thresholds of 
both 5 and 10\,GeV, with the asymmetric Guassian fit to P1 represented by the solid black curve.  Panel c shows 
the phase-folded VERITAS data from the location of the Geminga pulsar in units of counts per bin, and the average 
number of counts is indicated by the horizontal dashed line.  The vertical dashed lines in all three panels 
represent the phase regions for P1 and P2.  Figure from~\cite{andrew2015}.}
\label{fig:geminga_phaseograms}
\end{figure}

The VERITAS observational campaign on Geminga has resulted in the accumulation of a total of 71.6 hr of 
quality-selected data.
Event 
arrival times are barycentered and phase-folded with \texttt{Tempo2}~\cite{2006MNRAS.369..655H} using an 
{\it XMM-Newton} timing solution (E. Gotthelf 2014, private communication) for data obtained before the launch of 
the {\it Fermi}-LAT.  A publically available {\it Fermi}-LAT ephemeris from a webpage maintained by 
M. Kerr\footnote{\detokenize{www.slac.stanford.edu/~kerrm/fermi_pulsar_timing/}} is used for phase-folding all 
other VERITAS data.  The phase-folded Geminga pulsar data recorded by the {\it Fermi}-LAT shows two emission peaks 
(P1 and P2), which are used to define phase regions of expected signal for the VHE gamma-ray data taken by 
VERITAS. 
 The peaks P1 and P2 are both fitted with asymmetric Gaussian functions above $5$ and $10$\,GeV, respectively, 
which 
allows measuring the widths of the pulses at the highest energies possible in the {\it Fermi}-LAT band.  The two 
expected VHE signal regions are defined \emph{a priori} as the $\pm1\sigma$ regions for P1 and P2 ([0.072--0.125] 
and 
[0.575--0.617], respectively), and the background region is defined as [0.7--1.0]~\cite{andrew2015}.  The 
phase-folded {\it Fermi}-LAT and VERITAS light curves are shown in Figure~\ref{fig:geminga_phaseograms}.

\begin{figure}[H]
\center
\includegraphics[width=6.0in]{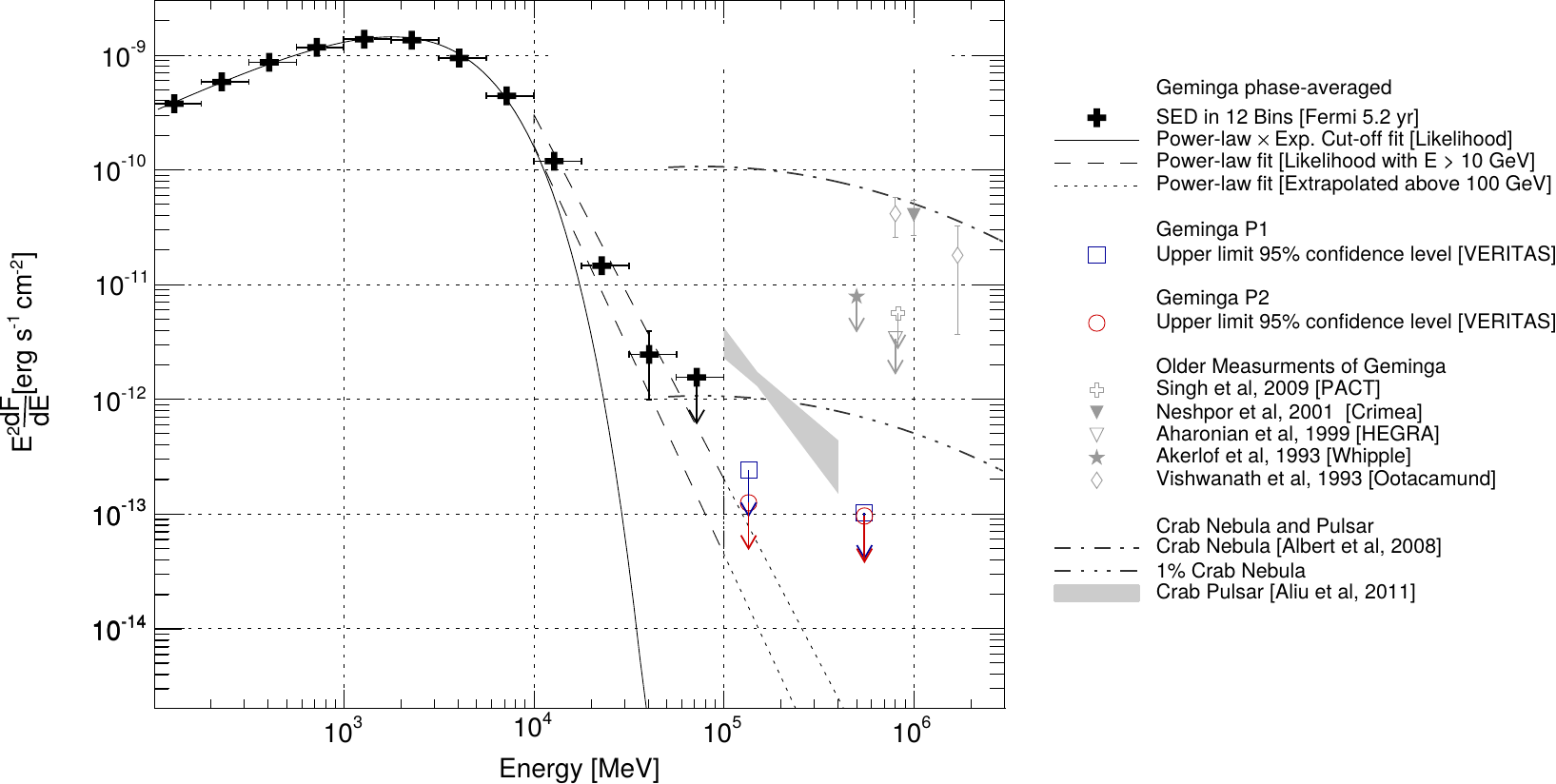}
\caption{Phase-averaged SED for the Geminga pulsar derived from 5.2\,yr of {\it Fermi}-LAT data with VHE flux ULs 
(open squares and circles) for the P1 and P2 phase regions computed from the VERITAS data.  The energy thresholds 
for the two sets of VHE flux ULs are 135 and 550\, GeV.  For a full description of the {\it Fermi}-LAT 
data analysis, see the recent VERITAS collaboration publication on the Geminga pulsar~\cite{andrew2015}.}
\label{fig:sed}
\end{figure}

The test for a pulsed signal using the $\pm1\sigma$ regions defined above reveals no evidence pulsed 
emission from the Geminga pulsar in the VERITAS data.  To more generally test for periodicity in the phase-folded 
VERITAS data, the {\it H}-Test~\cite{dejager1989} is employed.  The computed {\it H} statistic is $1.8$, 
corresponding to a random chance probability of 0.49, indicating that phase-folded VERITAS data is consistent with 
a random distribution.  Upper limits (ULs) on excess counts for the P1 and P2 phase regions are computed at the 
95\% confidence level (CL) using the method of~\cite{helene1983}.  These limits are converted to integral flux ULs 
above 135 GeV of $4.0 \times 10^{-13}\,\textrm{cm}^{-2}\,\textrm{s}^{-1}$ for P1 and $1.7 \times 
10^{-13}\,\textrm{cm}^{-2}\,\textrm{s}^{-1}$ for P2.  These limits are shown along with a phase-averaged {\it 
Fermi}-LAT SED for the Geminga pulsar in Figure~\ref{fig:sed}.  For a more complete explanation of this analysis 
and the results, see the recent VERITAS publication on the Geminga pulsar~\cite{andrew2015}.

\pagebreak
\subsection{PSR J1023+0038}

\begin{wrapfigure}{r}{0.5\textwidth}
\begin{center}
  \includegraphics[width=0.48\textwidth]{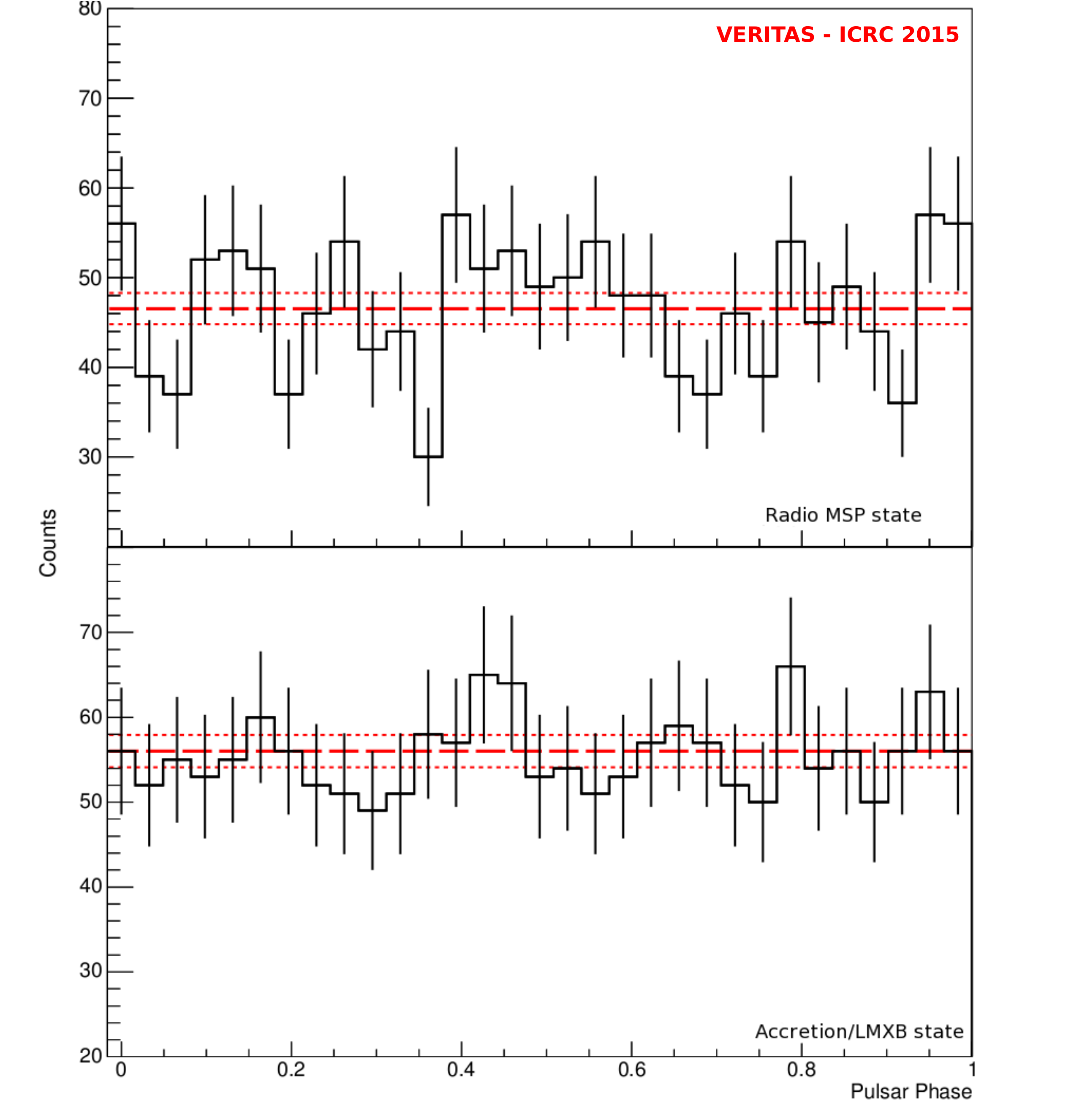}
\end{center}
\caption{Phase-folded light curves of events recorded by VERITAS from the direction of PSR J1023+0038.  One 
rotation of the pulsar is shown.  The top panel shows the data collected during the radio MSP state of the system, 
while the bottom panel shows the data collected during the accretion/LMXB state.  The solid and dashed lines 
represent the average and error on the average number of counts, respectively.}
\label{fig:j1023_phaseograms}
\end{wrapfigure}

\newcolumntype{C}{>{\centering\arraybackslash}X}
\begin{table}[h]
\scriptsize
\centering
\begin{tabularx}{0.7\textwidth}{CCCC}
  \hline
  J1023 State & H statistic & $2\sigma$ pulsed VHE flux UL (m$^{-2}$ s$^{-1}$) & $3\sigma$ pulsed VHE flux UL 
(m$^{-2}$ s$^{-1}$) \\ \hline
  Radio MSP & $0.28$  & $6.16 \times 10^{-9}$ & $9.73 \times 10^{-9}$     \\ 
   Accretion/LMXB & $0.18$ & $1.12 \times 10^{-8}$  & $1.97 \times 10^{-8}$ \\ 
\hline
\end{tabularx}
\caption{\normalfont{{\it H} statistics and integral pulsed VHE flux upper limits $>\,166$\,GeV for both the radio 
MSP and accrection/LMXB states of PSR J1023+0038.}}
\label{tab:pulsedfluxtable}
\end{table}

The search for pulsed VHE emission was split into two parts as VERITAS data were taken both before and after 
the 
2013 June disappearance of the radio pulsar (henceforth these two states of the system are referred to as the 
`radio MSP' and `accretion/LMXB' states).  The VERITAS live time accumulated on the target is 18.1\,hr 
for the radio MSP state and 8.2\,hr for the accretion/LMXB state. The data span two configurations of the VERITAS 
array: before and after the upgrade of the camera photomulitplier 
tubes in the Summer of 2012~\cite{kieda2013}.  Events are phase-folded in 
\texttt{Tempo2} using a Jodrell Bank radio ephemeris (A. Archibald, private 
communication), though the location of phase zero is unknown, precluding the possibility of 
defining signal regions in the phaseogram.  Therefore only the {\it H}-Test is used to test for the 
presence of a periodic signal in the data.  The {\it H}-Test 
does not reveal any evidence for periodicity in the VHE phaseograms, which are shown in 
Figure~\ref{fig:j1023_phaseograms}.  The {\it H} statistics are used to compute 2 and $3\sigma$ upper limits via 
the method of~\cite{1994ApJ...436..239D} above an energy threshold of 166\,GeV assuming a duty cycle of 10\% 
and 
Gaussian pulse shapes.  The computed {\it H} statistics and pulsed flux ULs are shown in 
Table~\ref{tab:pulsedfluxtable}.  A search for steady emission $> 300$\,GeV in the data is also performed, giving 
significances of $-0.8\sigma$ and $0.2\sigma$ for the radio MSP and accretion/LMXB states, respectively.  
Upper limits on a steady emission component are computed at the 95\% CL for both states assuming a photon 
index of $\Gamma = 2.5$ using the Rolke method~\cite{rolke2005}.  
The computed steady flux 95\% CL ULs are $8.1 \times 10^{-9}\,\textrm{m}^{-2}\,\textrm{s}^{-1}$ and $9.6 \times 
10^{-9}\,\textrm{m}^{-2}\,\textrm{s}^{-1}$ for the radio MSP and accretion/LMXB states, respectively.  
These results are part of a VERITAS collaboration publication that is currently in preparation.

\section{Future Plans}

Pulsar science with VERITAS is an ongoing effort and will continue in the coming years.  There are currently 19 
pulsar locations for which VERITAS data exist, including coverage of the top ten {\it Fermi}-LAT detected 
pulsars in the Northern Hemisphere when ranked in spin-down power divided by distance squared (for details, see 
Archer, A., these proceedings)~\cite{archer_pulsars}.  The analyses of 
these archival data sets is currently underway and will represent the first time that any of these pulsars 
have been probed for possible pulsed VHE emission.

Due to the good coverage of the young pulsars in archival VERITAS data, VERITAS is beginning to target 
a different promising population of pulsars to probe for VHE emission: millisecond pulsars.  Because of the 
rapid rotation of MSPs (spin periods reaching down to $\sim$1\,ms), these pulsars have much smaller light 
cylinder radii, implying  more compact magnetospheres.  The compactness of their magnetospheres 
suggests significantly different physical parameters of the emission mechanism compared to the young pulsar 
population.  Furthermore, though less luminous, their HE gamma-ray producion efficiencies have been found to be 
comparable or greater than those of the young pulsars~\cite{2013ApJS..208...17A}.  For these reasons, 
targeting MSPs in addition to young pulsars should provide complementary benefits for constraining models 
attempting to explain their gamma-ray emission.           

\section*{Acknowledgements}
This research is supported by grants from the U.S. Department of Energy Office of Science, 
the U.S. National Science Foundation and the Smithsonian Institution, by NSERC in Canada, by 
Science Foundation Ireland (SFI 10/RFP/AST2748). We acknowledge the excellent
work of the technical support staff at the Fred Lawrence Whipple Observatory and at the collaborating 
institutions in the construction and operation of the instrument.  The VERITAS Collaboration is grateful to Trevor 
Weekes for his seminal contributions and leadership in the field of VHE gamma-ray astrophysics, which made this 
study possible.


\begin{thebibliography}{99}
\bibliographystyle{ieeetr}
\scriptsize
\bibitem{1986ApJ...300..522C} Cheng, K.~S., Ho, C., \& Ruderman, M., \emph{Energetic Radiation from Rapidly 
Spinning Pulsars. II. VELA and Crab}, \emph{ApJ}, {\bf 300}, 522 (1986)
\bibitem{2013ApJS..208...17A} Abdo, A.~A., Ajello, M., Allafort, A., et al., \emph{The Second Fermi Large Area 
Telescope Catalog of Gamma-Ray Pulsars}, \emph{ApJS}, {\bf 208}, 17 (2013)
\bibitem{2009ApJ...696.1084A} Abdo, A.~A., Ackermann, M., Atwood, W.~B., et al., \emph{Fermi Large Area Telescope 
Observations of the Vela Pulsar}, \emph{ApJ}, {\bf 696}, 1084 (2009)
\bibitem{gajdus} Gajdus, M., \emph{Pulsations from the Vela pulsar down to 30 GeV with H.E.S.S. II}, in 
proceedings of \emph{34th International Cosmic Ray Conference} (2015)
\bibitem{leung2014} Leung, G.~C.~K., Takata, J., Ng, C.~W., et al., \emph{Fermi-LAT Detection of Pulsed Gamma-Rays 
above 50 GeV from the Vela Pulsar}, \emph{ApJL}, {\bf 797}, L13 (2014)
\bibitem{aleksic2011b} Aleksi{\'c}, J., Alvarez, E.~A., Antonelli, L.~A., et al., \emph{Observations of the Crab 
Pulsar between 25 and 100 GeV with the MAGIC I Telescope}, \emph{ApJ}, {\bf 742}, 43 (2011)
\bibitem{aliu2011} Aliu, E., Arlen, T., et al., \emph{Detection of Pulsed Gamma Rays Above 
100 GeV from the Crab Pulsar}, \emph{Science}, {\bf 334}, 69 (2011)
\bibitem{lyutikovetal2012} Lyutikov, M., Otte, N., \& McCann, A., \emph{The Very High Energy Emission from Pulsars: 
A Case for Inverse Compton Scattering}, \emph{ApJ}, {\bf 754}, 33 (2012)
\bibitem{du2012} Du, Y.~J., Qiao, G.~J., \& Wang, W., \emph{Radio-to-TeV Phase-resolved Emission from the Crab 
Pulsar: The Annular Gap Model}, \emph{ApJ}, {\bf 748}, 84 (2012)
\bibitem{lyutikov2012} Lyutikov, M., \emph{The {$\gamma$}-Ray Spectrum of Geminga and the Inverse Compton Model of 
Pulsar High-energy Emission}, \emph{ApJ}, {\bf 757}, 88 (2012)
\bibitem{aharonian2012} Aharonian, F.~A., Bogovalov, S.~V., \& Khangulyan, D., \emph{Abrupt acceleration of a 
`cold' ultrarelativistic wind from the Crab pulsar}, \emph{Nature}, {\bf 482}, 507 (2012)
\bibitem{petri2012} P{\'e}tri, J., \emph{High-energy emission from the pulsar striped wind: a synchrotron model for 
gamma-ray pulsars}, \emph{MNRAS}, {\bf 424}, 2023 (2012)
\bibitem{mocholpetri2015} Mochol, I., \& P\'etri, J., \emph{Very high energy emission as a probe of relativistic 
magnetic reconnection in pulsar winds}, \emph{MNRAS}, {\bf 449}, L51 (2015)
\bibitem{caraveo1996} Caraveo, P.~A., Bignami, G.~F., Mignani, R., \& Taff, L.~G., \emph{Parallax Observations with 
the Hubble Space Telescope Yield the Distance to Geminga}, \emph{ApJL}, {\bf 461}, L91 (1996)
\bibitem{faherty2007} Faherty, J., Walter, F.~M., \& Anderson, J., \emph{The trigonometric parallax of the neutron 
star Geminga}, \emph{Ap\&SS}, {\bf 308}, 225 (2007)
\bibitem{bignamicaraveo1996} Bignami, G.~F., \& Caraveo, P.~A., \emph{Geminga: Its Phenomenology, Its Fraternity, 
and Its Physics}, \emph{ARA\&A}, {\bf 34}, 331 (1996)
\bibitem{1992Natur.357..306B} Bertsch, D.~L., Brazier, K.~T.~S., Fichtel, C.~E., et al., \emph{Geminga: new period, 
old {$\gamma$}-rays}, \emph{Nature}, {\bf 357}, 306 (1992)
\bibitem{fichtel1975} Fichtel, C.~E., Hartman, R.~C., Kniffen, D.~A., et al., \emph{High-energy gamma-ray results 
from the second small astronomy satellite}, \emph{ApJ}, {\bf 198}, 163 (1975)
\bibitem{bennett1977} Bennett, K., Lichti, G.~G., Bignami, G.~F., et al., \emph{COS-B observations of 
localised high-energy gamma-ray emission from the anticentre region of the galactic disc}, \emph{A\&A}, {\bf 56}, 
469 (1977)
\bibitem{kargaltsev2005} Kargaltsev, O.~Y., Pavlov, G.~G., Zavlin, V.~E., \& Romani, R.~W., \emph{Ultraviolet, 
X-Ray, and Optical Radiation from the Geminga Pulsar}, \emph{ApJ}, {\bf 625}, 307 (2005)
\bibitem{halpernholt1992} Halpern, J.~P., \& Holt, S.~S., \emph{Discovery of soft X-ray pulsations from the 
gamma-ray source Geminga}, \emph{Nature}, {\bf 357}, 222 (1992)
\bibitem{deller2012} Deller, A.~T., Archibald, A.~M., Brisken, W.~F., et al., \emph{A Parallax Distance and Mass 
Estimate for the Transitional Millisecond Pulsar System J1023+0038}, \emph{ApJ}L, {\bf 756}, L25 (2012)
\bibitem{archibald2009} Archibald, A.~M., Stairs, I.~H., Ransom, S.~M., et al., \emph{A Radio Pulsar/X-ray Binary 
Link}, \emph{Science}, {\bf 324}, 1411 (2009)
\bibitem{2005AJ....130..759T} Thorstensen, J.~R., \& Armstrong, E., \emph{Is FIRST J102347.6+003841 Really a 
Cataclysmic Binary?}, \emph{AJ}, {\bf 130}, 759 (2005)
\bibitem{2004MNRAS.351.1015W} Woudt, P.~A., Warner, B., \& Pretorius, M.~L., \emph{High-speed photometry of faint 
cataclysmic variables - IV. V356 Aql, Aqr1, FIRST J1023+0038, H{$\alpha$} 0242-2802, GI Mon, AO Oct, V972 Oph, SDSS 
0155+00, SDSS 0233+00, SDSS 1240-01, SDSS 1556-00, SDSS 2050-05, FH Ser}, \emph{MNRAS}, {\bf 351}, 1015 (2004)
\bibitem{2013ATel.5513....1S} Stappers, B.~W., Archibald, A., Bassa, C., et al., \emph{State-change in the 
``transition'' binary millisecond pulsar J1023+0038}, \emph{The Astronomer's Telegram}, {\bf 5513}, 1 (2013)
\bibitem{2013ATel.5514....1H} Halpern, J.~P., Gaidos, E., Sheffield, A., Price-Whelan, A.~M., \& Bogdanov, S., 
\emph{Optical Observations of the Binary MSP J1023+0038 in a New Accreting State}, \emph{The Astronomer's 
Telegram}, {\bf 5514}, 1 (2013)
\bibitem{alpar1982} Alpar, M.~A., Cheng, A.~F., Ruderman, M.~A., \& Shaham, J., \emph{A new class of radio 
pulsars}, \emph{Nature}, {\bf 300}, 728 (1982)
\bibitem{2008AIPC.1085..657H} Holder, J., Acciari, V.~A., Aliu, E., et al., \emph{Status of the VERITAS 
Observatory}, \emph{American Institute of Physics Conference Series}, {\bf 1085}, 657 (2008)
\bibitem{2006MNRAS.369..655H} Hobbs, G.~B., Edwards, R.~T., \& Manchester, R.~N., \emph{TEMPO2, a new 
pulsar-timing package - I. An overview}, \emph{MNRAS}, {\bf 369}, 655 (2006)
\bibitem{andrew2015} Aliu, E., Archambault, S., Archer, A., et al., \emph{A Search for Pulsations from Geminga 
above 100 GeV with VERITAS}, \emph{ApJ}, {\bf 800}, 61 (2015)
\bibitem{dejager1989} de Jager, O.~C., Raubenheimer, B.~C., \& Swanepoel, J.~W.~H., \emph{A poweful test for weak 
periodic signals with unknown light curve shape in sparse data}, \emph{A\&A}, {\bf 221}, 180 (1989)
\bibitem{helene1983} Helene, O., \emph{Upper limit of peak area}, \emph{Nuclear Instruments and Methods in Physics 
Research}, {\bf 212}, 319 (1983)
\bibitem{kieda2013} Kieda, D., et al., \emph{Status of the VERITAS Upgrade}, in proceedings of \emph{32nd 
International Cosmic Ray Conference}, {\bf 9}, 14 (2011)
\bibitem{1994ApJ...436..239D} de Jager, O.~C., \emph{On periodicity tests and flux limit calculations for gamma-ray 
pulsars}, \emph{ApJ}, {\bf 436}, 239 (1994)
\bibitem{rolke2005} Rolke, W.~A., \& L{\'o}pez, A.~M., \emph{Confidence intervals and upper bounds for small 
signals in the presence of background noise}, \emph{Nuclear Instruments and Methods in Physics Research A}, {\bf 
458}, 745 (2005)
\bibitem{archer_pulsars} Archer, A., \emph{Search for pulsed emission in archival VERITAS data}, in proceedings of 
\emph{34th International Cosmic Ray Conference} (2015) 

\end{thebibliography}

\end{document}